\documentclass[10pt]{article}
\usepackage{graphicx}
\usepackage{indentfirst}
\usepackage{multicol}
\usepackage{amsfonts}
\usepackage{amsmath}

\begin{document}

\centerline{\LARGE Unambiguous unitary quantum channels}

\medskip

\centerline{\large Shengjun Wu and Xuemei Chen}

\centerline{\it Hefei National Laboratory for Physical Sciences at
Microscale and Department of Modern Physics}

\centerline{\it University of Science and Technology of China, Hefei,
Anhui 230026, P. R. China }

\smallskip

\centerline{\today}

\begin{abstract}

Unambiguous unitary maps and unambiguous unitary quantum channels
are introduced and some of their properties are derived. These
properties ensure certain simple form for the measurements involved
in realizing an unambiguous unitary quantum channel. Error
correction and unambiguous error correction with nonzero probability
are discussed in terms of unambiguous unitary quantum channels. We
not only re-derive the well-known condition for a set of errors to
be correctable with certainty, but also obtain a necessary and
sufficient condition for the errors caused by a noisy channel to be
correctable with any nonzero probability. Dense coding with a
partially entangled state can also be viewed as an unambiguous
unitary quantum channel when all messages are required to be
transmitted with equal probability of success, the maximal
achievable probability of success is derived and the optimum
protocol is also obtained.

\end{abstract}


\section{Introduction}
\nopagebreak

Quantum teleportation \cite{bbcjpw} gives us an example on how a
maximally entangled state shared between Alice and Bob can be used
to transfer an unknown state with both perfect fidelity and
certainty. Unambiguous teleportation \cite{horodeckimpr,bhm} with
shared partially entangled state as a resource can transmit an unknown state with
perfect fidelity although the probability of success is less than
unity. Quantum errors occur when the quantum states go through a
noisy quantum channel, but some errors can be corrected by quantum
error correction \cite{shor95, steane96}. If the errors are correctable, after quantum error
correction, the quantum state can be transformed back
into the original state with fidelity 1.

There are many other examples that have the same property:
an unknown state in a certain Hilbert space (or subspace) needs to be
transmitted with perfect fidelity, although the probability of
success could be less than 1. Both the sender and the receiver should know
when the unknown state is successfully transmitted, namely the unknown state
should be transmitted with perfect fidelity unambiguously.
This kind of process can be conveniently discussed with the concept of unambiguous unitary
quantum channels that are introduced in this paper.

The structure of this paper is arranged as follows.
In Sec. \ref{UUQC}, we introduce unambiguous unitary maps and unambiguous unitary quantum channels,
with some useful properties derived. In particular the equivalence of an unambiguous unitary quantum channel
and a uniformly entangled state is established.

Unambiguous teleportation with partially entangled state as a resource is considered
as an unambiguous unitary quantum channel and discussed in Sec. \ref{contele}.
When Alice and Bob share an (pure or mixed) entangled state, the operation elements of certain measurement involved in
their general LOCC operations for unambiguous
teleportation
can be chosen as rank-one operators.

In Sec. \ref{qec}, the combined operation
of a noisy quantum channel and the error correction is viewed as an unambiguous unitary quantum channel.
The well-known condition for correctable
errors are easily re-derived using the properties of the unambiguous unitary quantum channels.
We also discuss unambiguous error correction, and obtain a necessary and sufficient condition for
the errors to be corrected with any nonzero probability.

In Sec. \ref{cdc}, unambiguous dense coding with the requirement that all messages are to be sent through with equal probability of success
is considered as an unambiguous unitary quantum channel. An upper bound on the probability of success and the protocol to achieve it are both obtained.

\section{Unambiguous unitary quantum channels}  ~\label{UUQC}
\nopagebreak

\subsection{Unambiguous communication of an unknown state}  ~\label{UUQCcqic}
\nopagebreak

Unambiguous (conclusive) teleportation is the process that occurs
when Alice wants to send Bob an unknown state with perfect fidelity
using two kinds of resources: a partially entangled state and a
classical channel. The unknown state should be sent to Bob with
perfect fidelity when their protocol succeeds, and they should know
when their protocol succeeds.

There are other processes that preserve an unknown state with perfect fidelity, for example,
the combined operation of quantum error correction and a noisy quantum channel.
They are all quantum operations that preserve an arbitrary state in a certain Hilbert space (subspace)
with perfect fidelity (the probability of success could be less than unity). This kind of operation
is called an unambiguous unitary quantum channel, with detailed discussion given in the next two subsections.
In other words, an unambiguous unitary quantum channel is a natural generalization of the unambiguous teleportation,
without mentioning how the operation is implemented and what resources are used.

In order to implement an unambiguous unitary quantum channel that involves more than one observer, some shared resources are needed.
Throughout this paper when we mention shared resources, we have the following 4 kinds in mind: (1) (perfect or noisy) quantum channels,
(2) entangled states, (3) classical channels, (4) classical random bits. The first two are referred to as the shared
quantum resources, and the latter two as the shared classical resources.

\subsection{Unambiguous unitary maps} ~\label{UUM}
\nopagebreak

The following abbreviations
\begin{equation}
\mathcal{H}_{12} = \mathcal{H}_{1} \otimes \mathcal{H}_{2} ,  \quad
\hat{\mathcal{H}}_{1} = \mathcal{H}_{1} \otimes \mathcal{H}_{1}^{\dag} , \quad
\mathcal{L}_{21} = \mathcal{H}_2 \otimes \mathcal{H}_1^{\dag}
\end{equation}
are adopted for the tensor product of two Hilbert spaces,
the space of operators, and the space of linear maps from $\mathcal{H}_1$ to $\mathcal{H}_2$.
The notation $\hat{\mathcal{H}}_{1}$ is usually denoted by $\mathcal{B}(\mathcal{H}_{1})$ in the literature.

We consider a very general quantum operator $\Omega$ :
$\mathcal{H}_{I} \rightarrow \mathcal{H}_{O}$, which is a map from
the input Hilbert space $\mathcal{H}_{I}$ that is the product space
of the Hilbert spaces $\mathcal{H}_i$ ($i=1,3,\cdots$) of the input
systems $\mathcal{S}_i$ ($i=1,3,\cdots$), to the output Hilbert
space $\mathcal{H}_{O}$ that is the product space of the Hilbert
spaces $\mathcal{H}_i$ ($i=2,4,\cdots$) of the output systems
$\mathcal{S}_i$ ($i=2,4,\cdots$).  Let $\mathcal{H}_{e_1}$ denote
the Hilbert space of the combined system $e_1$ that includes all the
input systems except system $1$, $\mathcal{H}_{I}=\mathcal{H}_{1}
\otimes \mathcal{H}_{e_1}$, and $\mathcal{H}_{e_2}$ denote the
Hilbert space of the combined system $e_2$ that includes the output
systems except system $2$, $\mathcal{H}_{O}=\mathcal{H}_{2} \otimes
\mathcal{H}_{e_2}$, and let $\mathcal{H}_{e}$ denote their product,
$\mathcal{H}_{e}=\mathcal{H}_{e_1} \otimes \mathcal{H}_{e_2}$. Using
the atemporal diagram approach \cite{griffiths2005}, such an operator (map)
can be represented by a square with directed lines (legs) attached
(see FIG. 1), a leg points towards the box if it is associated with
an input system, or points against the box if it is associated with
an output system.
\begin{center}
\setlength{\unitlength}{0.05in}
\begin{picture}(43,8)(0,-5)
\thicklines

\put(7,-3){\vector(1,0){3}}
\put(9,-2.5){\makebox(0,0)[b]{\small $1$}}

\put(10,-5){\framebox(6,4)[]{$\Omega$}}
\put(11,2){\vector(0,-1){3}}
\put(12,-1){\vector(0,1){3}}
\put(15,-1){\vector(0,1){3}}
\put(13.5,0){\makebox(0,0)[c]{\small $\cdot \cdot$}}
\put(16,-3){\vector(1,0){3}}
\put(18,-2.5){\makebox(0,0)[b]{\small $2$}}
\put(10.5,2.5){\makebox(0,0)[b]{\footnotesize $3$}}
\put(12,2.5){\makebox(0,0)[b]{\footnotesize $4$}}
\put(14,2.5){\makebox(0,0)[b]{\footnotesize $\cdots$}}

\label{pictpqc001}
\end{picture}

\nopagebreak
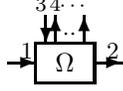
\noindent{FIG. 1.  A general operator as a many-leg box }
\end{center}

{\bf Definition 1. } An operator $\Omega$ is called an unambiguous unitary map (UUM) from a $d$-dimensional subspace $\mathcal{H}^s_1$ of $\mathcal{H}_1$
to a $d$-dimensional subspace $\mathcal{H}^s_2$ of $\mathcal{H}_2$,
if and only if there exists an operator $U$, a unitary map from $\mathcal{H}^s_1$ to $\mathcal{H}^s_2$, such that
for any pure state $\left| \psi \right\rangle \in \mathcal{H}^s_1$,
we have
\begin{equation}
P_2^s Tr_{e_2} \left\{ \Omega \left| \psi \right\rangle \left\langle \psi \right| \Omega^{\dag} \right\} P_2^s  = p \; U \left| \psi \right\rangle \left\langle \psi \right| U^{\dag}
\label{pqcdef}
\end{equation}
where $P_2^s$ is the projector onto the subspace $\mathcal{H}_2^s$
and $p>0$.
This definition is explicitly illustrated by FIG. 2.
\begin{center}
\setlength{\unitlength}{0.05in}
\begin{picture}(48,13)(7,-5)
\thicklines

\put(7,-3){\vector(1,0){3}}
\put(9,-2.5){\makebox(0,0)[b]{\small $1$}}

\put(10,-5){\framebox(6,4)[]{$\Omega$}}
\put(11,2){\vector(0,-1){3}}
\put(12,-1){\vector(0,1){3}}
\put(15,-1){\vector(0,1){3}}
\put(13.5,0){\makebox(0,0)[c]{\small $\cdot \cdot$}}
\put(16,-3){\vector(1,0){3}}
\put(18,-2.5){\makebox(0,0)[b]{\small $2$}}
\put(19,-5){\framebox(4,4)[]{$P_2^s$}}
\put(23,-3){\vector(1,0){3}}
\put(25,-2.5){\makebox(0,0)[b]{\small $2$}}

\put(10,4){\vector(-1,0){3}}
\put(9,4.5){\makebox(0,0)[b]{\small $1$}}
\put(10,2){\framebox(6,4)[]{$\Omega^\dag$}}
\put(19,4){\vector(-1,0){3}}
\put(18,4.5){\makebox(0,0)[b]{\small $2$}}
\put(19,2){\framebox(4,4)[]{$P_2^s$}}
\put(26,4){\vector(-1,0){3}}
\put(25,4.5){\makebox(0,0)[b]{\small $2$}}

\put(3,-5){\framebox(4,4)[]{$\left| \psi \right>$}}
\put(3,2){\framebox(4,4)[]{$\left< \psi \right|$}}

\put(28,-0.5){\makebox(0,0)[b]{$\propto$}}

\put(33,0){\vector(-1,0){3}}
\put(32,0.5){\makebox(0,0)[b]{\small $2$}}
\put(33,-2){\framebox(4,4)[]{$U$}}
\put(40,0){\vector(-1,0){3}}
\put(39,0.5){\makebox(0,0)[b]{\small $1$}}
\put(40,-2){\framebox(4,4)[]{$\left| \psi \right>$}}
\put(45,-2){\framebox(4,4)[]{$\left< \psi \right|$}}
\put(52,0){\vector(-1,0){3}}
\put(51,0.5){\makebox(0,0)[b]{\small $1$}}
\put(52,-2){\framebox(4,4)[]{$U^\dag$}}
\put(59,0){\vector(-1,0){3}}
\put(58,0.5){\makebox(0,0)[b]{\small $2$}}

\label{pictpqc002}
\end{picture}

\nopagebreak
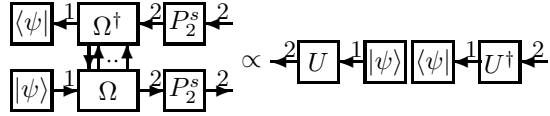
\noindent{FIG. 2.  An unambiguous unitary map $\Omega_{\left( U, p \right)}$ acting on any pure state $\left| \psi \right> \in \mathcal{H}_1^s$}
\end{center}

The positive number $p$ in (\ref{pqcdef}) is not explicitly required to be independent of the input state $\left| \psi \right>$ in the definition
of an UUM; however as we shall see in proposition 1, it is actually always independent of the input state, a fact {\sl implied} by the definition of an UUM.
The number $p$ is the probability that $\Omega$ can act like $U$: taking any input state in $\mathcal{H}_1^s$ to a state
in $\mathcal{H}_2^s$ according to what $U$ does.
We can check whether $\Omega$ mimics a unitary map successfully by projecting the final state into the subspace
$\mathcal{H}_2^s$, hence the word {\sl unambiguous} is used.
An operator $\Omega$ that is an unambiguous unitary map from $\mathcal{H}_1^s$ to $\mathcal{H}_2^s$ with
probability $p$ and corresponds to the unitary map $U$ can be conveniently denoted by
$\Omega = \Omega_{\left( U, p \right)}$, we may simply say
that $\Omega$ is a probabilistic (unitary map) $U$ with probability $p$.  The two related subspaces are uniquely specified by the unitary map $U$,
sometimes it is convenient to explicitly specify the dimension $d$ of the subspaces by $\Omega = \Omega_{\left( U, p , d \right)}$.

{\bf Proposition 1. } An unambiguous unitary map has no preference on the input, i.e., the probability $p$ in (\ref{pqcdef})
is independent of the input state $\left| \psi \right\rangle $ ($\in \mathcal{H}_1^s$).

{\bf Proposition 2. }
An operator $\Omega$ is an unambiguous unitary map that mimics $U$
from a subspace $\mathcal{H}^s_1$ ($ \subset \mathcal{H}_1$) to a subspace $\mathcal{H}^s_2$ ($ \subset \mathcal{H}_2$) with probability $p>0$
(namely, $\Omega = \Omega_{\left( U, p \right)}$),
if and only if
\begin{equation}
P^s_2 \Omega P^s_1 = U \otimes \Theta  \; .   \label{proposition1it3}
\end{equation}
where $P^s_1$ and $P^s_2$ are projectors onto the subspaces $\mathcal{H}^s_1$ and $\mathcal{H}^s_2$ respectively, and $\Theta$,
an object with $m-2$ legs (legs $3,\cdots,m$), is related to the probability $p$ by
$p= Tr_{e_2} \left\{ \Theta \Theta^{\dag} \right\} $.
The proof of the two propositions will be given in the appendix.
\begin{center}
\setlength{\unitlength}{0.05in}
\begin{picture}(43,14)(0,-7.5)
\thicklines

\put(0,-3){\vector(1,0){3}}
\put(3,-5){\framebox(4,4)[]{$P_1^s$}}
\put(7,-3){\vector(1,0){3}}
\put(2,-2.5){\makebox(0,0)[b]{\small $1$}}
\put(9,-2.5){\makebox(0,0)[b]{\small $1$}}

\put(10,-5){\framebox(6,4)[]{$\Omega$}}
\put(11,2){\vector(0,-1){3}}
\put(12,-1){\vector(0,1){3}}
\put(15,-1){\vector(0,1){3}}
\put(13.5,0){\makebox(0,0)[c]{\small $\cdot \cdot$}}
\put(16,-3){\vector(1,0){3}}
\put(18,-2.5){\makebox(0,0)[b]{\small $2$}}
\put(19,-5){\framebox(4,4)[]{$P_2^s$}}
\put(23,-3){\vector(1,0){3}}
\put(25,-2.5){\makebox(0,0)[b]{\small $2$}}
\put(10.5,2.5){\makebox(0,0)[b]{\footnotesize $3$}}
\put(12,2.5){\makebox(0,0)[b]{\footnotesize $4$}}
\put(14,2.5){\makebox(0,0)[b]{\footnotesize $\cdots$}}

\put(29,-3){\makebox(0,0)[c]{$=$}}

\put(33,-6){\vector(1,0){3}}
\put(36,-8){\framebox(4,4)[]{$U$}}
\put(40,-6){\vector(1,0){3}}
\put(35,-5.5){\makebox(0,0)[b]{\small $1$}}
\put(42,-5.5){\makebox(0,0)[b]{\small $2$}}

\put(35,-2){\framebox(6,4)[]{$\Theta$}}
\put(36,5){\vector(0,-1){3}}
\put(37,2){\vector(0,1){3}}
\put(40,2){\vector(0,1){3}}
\put(38.5,3){\makebox(0,0)[c]{\small $\cdot \cdot$}}
\put(35.5,5.5){\makebox(0,0)[b]{\footnotesize $3$}}
\put(37,5.5){\makebox(0,0)[b]{\footnotesize $4$}}
\put(39,5.5){\makebox(0,0)[b]{\footnotesize $\cdots$}}

\label{pictpqc2}
\end{picture}

\nopagebreak
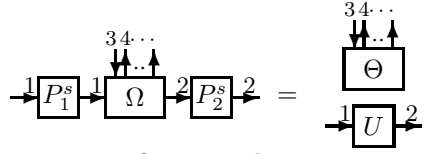
\noindent{FIG. 3.  The structure of an unambiguous unitary map $\Omega_{\left( U, p \right)}$}
\end{center}

It is implied by proposition 2 that {\it any unambiguous unitary map,
conditional on the subspaces of the two legs we are interested in,
must have all other legs detached.}
This result is also true if the unitary map $U$ is replaced by any reversible map in the definition of an UUM.
Since such a generalized map can always be turned into an unambiguous unitary map (with possibly less
probability) by performing additional physically realizable local operation on $\mathcal{H}_2^s$.

From (\ref{proposition1it3}), it is obvious that
there exists a unitary map $U$ from $\mathcal{H}^s_1$ to $\mathcal{H}^s_2$ such that
for any state $\left| \psi \right\rangle \in \mathcal{H}^s_1$
\begin{equation}
P^s_2 \Omega \left| \psi \right\rangle = (U\left| \psi \right\rangle) \otimes \Theta \label{proposition1it2}  \; .
\end{equation}
This can be viewed as an alternative definition of an UUM.

\subsection{Unambiguous unitary quantum channels} ~\label{UUQCsubs}
\nopagebreak

A quantum operation $\mathcal{E}$ from some input systems $\mathcal{S}_i$ ($i=1,3,\cdots$) to some output systems $\mathcal{S}_i$ ($i=2,4,\cdots$)
can be described as a map from the set of density operators for the input systems to the set of density operators for the output systems.
Since the density operators for the input (output) systems can be viewed as vectors in the Hilbert space $\hat{\mathcal{H}}_{I}$
($\hat{\mathcal{H}}_{O}$), $\mathcal{E}$ is a map $\mathcal{E}$: $\hat{\mathcal{H}}_{I} \rightarrow \hat{\mathcal{H}}_{O}$ from $\hat{\mathcal{H}}_{I}$
to $\hat{\mathcal{H}}_{O}$.
In the operator-sum representation, $\mathcal{E}$ can be described as
\begin{equation}
\mathcal{E} \left( \rho_{I} \right)  = \sum_k \Omega_k \rho_{I} \Omega_k^{\dag}  \label{operatorsum}
\end{equation}
where the operators $\{ \Omega_k \}$, which are maps from the input Hilbert space $\mathcal{H}_{I}$
to the output Hilbert space $\mathcal{H}_{O}$,
are known as operation elements (or Kraus operators) for the quantum operation $\mathcal{E}$.
The quantum operation $\mathcal{E}$ given in (\ref{operatorsum}) can be realized physically if and only if
\begin{equation}
\sum_k \Omega_k^{\dag} \Omega_k \leq I_{I}  \label{physical}
\end{equation}
where $I_I$ is the identity operator on $\mathcal{H}_I$.
In terms of the positive-operator-valued measure (POVM) formalism,
the operators $G_x = \Omega_k^{\dag} \Omega_k$ are known as the POVM elements \cite{NielsenChuang}.
For a given quantum operation $\mathcal{E}$, the set of operation elements $\{ \Omega_k \}$
in its operator-sum representation (\ref{operatorsum}) may not be unique.

Suppose $\mathcal{H}^s_1$ is a $d$-dimensional subspace  of $\mathcal{H}_1$, and $\mathcal{H}^s_2$ is a
$d$-dimensional subspace of $\mathcal{H}_2$.
Let $\hat{\mathcal{H}}_1^s$ ($\hat{\mathcal{H}}_2^s$) denote the space of operators whose supports and ranges both lie in
$\mathcal{H}_1^s$ ($\mathcal{H}_2^s$), in a less rigorous notation,
$\hat{\mathcal{H}}_1^s = \mathcal{H}^s_1 \otimes \mathcal{H}^{s\dag}_1$,
$\hat{\mathcal{H}}_2^s = \mathcal{H}^s_2 \otimes \mathcal{H}^{s\dag}_2$.

{\bf Definition 2.}
A quantum operation $\mathcal{E}$ is called an unambiguous unitary quantum channel (UUQC)
from $\hat{\mathcal{H}}_1^s$ to $\hat{\mathcal{H}}_2^s$ with probability $q$,
if and only if there exists a unitary channel $\mathcal{U}$ from $\hat{\mathcal{H}}_1^s$ to $\hat{\mathcal{H}}_2^s$ such that
for any density operator $\rho_{1}$ whose support lies in $\mathcal{H}^s_1$,
we have
\begin{equation}
P_2^s Tr_{e_2} \left\{ \mathcal{E} ( \rho_{1} \otimes I_{e_1} ) \right\} P_2^s  = q \; \mathcal{U} \left( \rho_1 \right)
\label{pqcdef2}
\end{equation}
with $q>0$, here $P_2^s$ is the projector onto the subspace $\mathcal{H}_2^s$, and $I_{e_1}$ is the identity operator
for the input systems except system $\mathcal{S}_1$.  Here $\mathcal{U}$ is the unitary quantum channel that takes any density operator
$\rho_{1}$ whose support lies in $\mathcal{H}^s_1$ to a density operator
\begin{equation}
\rho_2 \equiv \mathcal{U} \left( \rho_{1} \right)  = U \rho_{1} U^{\dag}  \label{mapU}
\end{equation}
whose support lies in $\mathcal{H}^s_2$.
Such a quantum operation $\mathcal{E}$ can be conveniently denoted by $\mathcal{E}_{\left( \mathcal{U}, q \right)}$
(or $\mathcal{E}_{\left(U, q \right)}$ when there is no confusion), and simply be called
a probabilistic $\mathcal{U}$ (with probability $q$). The notation $\mathcal{E}_{\left(U, q , d\right)}$
is also used to explicitly specify the dimension of the subspace $\mathcal{H}^s_1$ or $\mathcal{H}^s_2$.
\begin{center}
\setlength{\unitlength}{0.05in}
\begin{picture}(48,13)(3,-5)
\thicklines

\put(7,-3){\vector(1,0){3}}
\put(9,-2.5){\makebox(0,0)[b]{\small $1$}}

\put(10,-5){\framebox(6,4)[]{$\Omega_k$}}
\put(11,2){\vector(0,-1){3}}
\put(12,-1){\vector(0,1){3}}
\put(15,-1){\vector(0,1){3}}
\put(13.5,0){\makebox(0,0)[c]{\small $\cdot \cdot$}}
\put(16,-3){\vector(1,0){3}}
\put(18,-2.5){\makebox(0,0)[b]{\small $2$}}
\put(19,-5){\framebox(4,4)[]{$P_2^s$}}
\put(23,-3){\vector(1,0){3}}
\put(25,-2.5){\makebox(0,0)[b]{\small $2$}}

\put(10,4){\vector(-1,0){3}}
\put(9,4.5){\makebox(0,0)[b]{\small $1$}}
\put(10,2){\framebox(6,4)[]{$\Omega_k^\dag$}}
\put(19,4){\vector(-1,0){3}}
\put(18,4.5){\makebox(0,0)[b]{\small $2$}}
\put(19,2){\framebox(4,4)[]{$P_2^s$}}
\put(26,4){\vector(-1,0){3}}
\put(25,4.5){\makebox(0,0)[b]{\small $2$}}

\put(3,-4){\framebox(4,10)[]{$\rho_1$}}
\put(0,-0.5){\makebox(0,0)[b]{$\sum_k$}}

\put(28,-0.5){\makebox(0,0)[b]{$\propto$}}

\put(33,0){\vector(-1,0){3}}
\put(32,0.5){\makebox(0,0)[b]{\small $2$}}
\put(33,-2){\framebox(4,4)[]{$U$}}
\put(40,0){\vector(-1,0){3}}
\put(39,0.5){\makebox(0,0)[b]{\small $1$}}
\put(40,-2){\framebox(4,4)[]{$\rho_1$}}
\put(47,0){\vector(-1,0){3}}
\put(46,0.5){\makebox(0,0)[b]{\small $1$}}
\put(47,-2){\framebox(4,4)[]{$U^\dag$}}
\put(54,0){\vector(-1,0){3}}
\put(53,0.5){\makebox(0,0)[b]{\small $2$}}

\label{pictpqc005}
\end{picture}

\nopagebreak
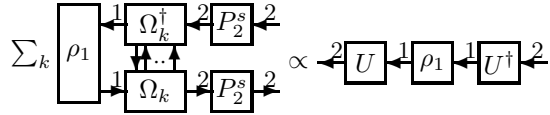
\noindent{FIG. 4.  Definition of an UUQC $\mathcal{E}_{\left( U, q \right)}$ in terms of its operation elements $\{ \Omega_k \}$}
\end{center}

Intuitively, $q$ is the probability that $\mathcal{E}$ can mimic the unitary channel $\mathcal{U}$ without any error,
and we can check whether $\mathcal{E}$ mimics the unitary channel $\mathcal{U}$ faithfully by checking whether the
result state is in the subspace $\mathcal{H}^s_2$, this is done by measuring $P^s_2$.
When $q=1$ in (\ref{pqcdef2}), the quantum operation $\mathcal{E}$ should be a trace-preserving operation,
and the support of
$Tr_{e_2} \left\{ \mathcal{E} ( \rho_{1} \otimes I_{e_1} ) \right\}$ should lie in $\mathcal{H}^s_2$
(namely, $P^s_2$ can be removed from (\ref{pqcdef2}) in this case). In general, the probability $q$ in (\ref{pqcdef2}) could be
less than $1$, due to the projector $P_2^s$, or a trace-decreasing $\mathcal{E}$, or both.

{\bf Proposition 3.}
If an UUQC $\mathcal{E}$ is a probabilistic $\mathcal{U}$ from $\hat{\mathcal{H}}_1^s$ to $\hat{\mathcal{H}}_2^s$
with probability $q$, $\mathcal{E}= \mathcal{E}_{\left(U, q \right)}$, and the set of operators $\{ \Omega_k \}$ is an arbitrary choice of
its operation elements, then each operator $\Omega_k$ must be an unambiguous unitary map $U$
from $\mathcal{H}^s_1$ to $\mathcal{H}^s_2$ with some probability $p_k \geq 0$, and $\sum_k p_k =q$.
(This can be easily proved from (\ref{operatorsum}) and (\ref{pqcdef2})).
Without loss of generality, suppose $p_k >0$ for $1\leq k \leq L$ and $p_k =0$ for $k >L$.
According to proposition 2, the operators $\Omega_k$ ($1\leq k \leq L$) that correspond to nonzero probabilities can be written as
\begin{equation}
P_2^s \Omega_k P_1^s = U \otimes \Theta_k   \label{omegakthetak}
\end{equation}
where $U$ is the map from $\mathcal{H}_1^s$ to $\mathcal{H}_2^s$, related to $\mathcal{U}$ by (\ref{mapU}), and $\Theta_k$ is a map
from $\mathcal{H}_{e_1}$ to $\mathcal{H}_{e_2}$, related to the probability $p_k$ by
\begin{equation}
p_k = Tr_{e_2} \left( \Theta_k^{\dag} \Theta_k \right) . \label{pk}
\end{equation}
$P_1^s$ and $P_2^s$ are projectors onto the subspaces $\mathcal{H}_1^s$ and $\mathcal{H}_2^s$ respectively.
\begin{center}
\setlength{\unitlength}{0.05in}
\begin{picture}(43,14)(0,-7.5)
\thicklines

\put(0,-3){\vector(1,0){3}}
\put(3,-5){\framebox(4,4)[]{$P_1^s$}}
\put(7,-3){\vector(1,0){3}}
\put(2,-2.5){\makebox(0,0)[b]{\small $1$}}
\put(9,-2.5){\makebox(0,0)[b]{\small $1$}}

\put(10,-5){\framebox(6,4)[]{$\Omega_k$}}
\put(11,2){\vector(0,-1){3}}
\put(12,-1){\vector(0,1){3}}
\put(15,-1){\vector(0,1){3}}
\put(13.5,0){\makebox(0,0)[c]{\small $\cdot \cdot$}}
\put(16,-3){\vector(1,0){3}}
\put(18,-2.5){\makebox(0,0)[b]{\small $2$}}
\put(19,-5){\framebox(4,4)[]{$P_2^s$}}
\put(23,-3){\vector(1,0){3}}
\put(25,-2.5){\makebox(0,0)[b]{\small $2$}}
\put(10.5,2.5){\makebox(0,0)[b]{\footnotesize $3$}}
\put(12,2.5){\makebox(0,0)[b]{\footnotesize $4$}}
\put(14,2.5){\makebox(0,0)[b]{\footnotesize $\cdots$}}

\put(29,-3){\makebox(0,0)[c]{$=$}}

\put(33,-6){\vector(1,0){3}}
\put(36,-8){\framebox(4,4)[]{$U$}}
\put(40,-6){\vector(1,0){3}}
\put(35,-5.5){\makebox(0,0)[b]{\small $1$}}
\put(42,-5.5){\makebox(0,0)[b]{\small $2$}}

\put(35,-2){\framebox(6,4)[]{$\Theta_k$}}
\put(36,5){\vector(0,-1){3}}
\put(37,2){\vector(0,1){3}}
\put(40,2){\vector(0,1){3}}
\put(38.5,3){\makebox(0,0)[c]{\small $\cdot \cdot$}}
\put(35.5,5.5){\makebox(0,0)[b]{\footnotesize $3$}}
\put(37,5.5){\makebox(0,0)[b]{\footnotesize $4$}}
\put(39,5.5){\makebox(0,0)[b]{\footnotesize $\cdots$}}

\label{pictpqc0006}
\end{picture}

\nopagebreak
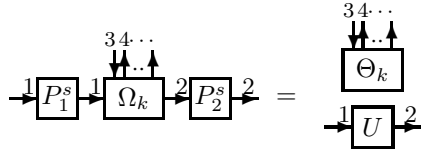
\noindent{FIG. 5.  An operation element $\Omega_k$ of $\mathcal{E}_{\left( U, q \right)}$}
\end{center}

Now we show that each operator $\Theta_k$ in (\ref{omegakthetak})
can be refined to a simple product form by refining the operation
elements $\{ \Omega_k \}$, which are unambiguous unitary maps with
nonzero probabilities. The operators $\Theta_k$ in
(\ref{omegakthetak}) are maps from the Hilbert space
$\mathcal{H}_{e_1}$ ($=\mathcal{H}_3 \otimes \mathcal{H}_5 \otimes
\cdots$) of the input systems (except system 1) to the Hilbert space
$\mathcal{H}_{e_2}$ ($=\mathcal{H}_4 \otimes \mathcal{H}_6 \otimes
\cdots$) of the output systems (except system 2).  Let $\{\left|
i_l\right>\}$ ($l=3,4,\cdots$) be an arbitrary basis of
$\mathcal{H}_l$ for system $\mathcal{S}_l$. We define the following
refinement of the operation elements,
\begin{equation}
\Omega_{k i_3 i_4 i_5 \cdots}  \equiv \left( P_2^s \otimes \left| i_4 \right> \left< i_4 \right| \otimes \cdots \right) \Omega_k
     \cdot \left( P_1^s \otimes \left| i_3 \right> \left< i_3 \right| \otimes \cdots \right)  .  \label{omegarefine}
\end{equation}
From (\ref{omegakthetak}) and (\ref{omegarefine}) we have
\begin{equation}
\Omega_{k i_3 i_4 i_5 \cdots} =  \omega_{k i_3 i_4 i_5 \cdots}  U \otimes \left( \left| i_4 \right>
\otimes \left| i_6 \right> \otimes \cdots \right)
 \otimes \left( \left< i_3 \right| \otimes \left< i_5 \right| \otimes \cdots \right)
 \label{omegakiii}
\end{equation}
where $\omega_{k i_3 i_4 i_5 \cdots}$ are complex numbers given by
\begin{equation}
\omega_{k i_3 i_4 i_5 \cdots} =  \left( \left< i_4 \right| \otimes \left< i_6 \right| \otimes \cdots \right) \Theta_k \left( \left| i_3 \right> \otimes \left| i_5 \right> \otimes \cdots \right).
\end{equation}
From (\ref{omegakiii}) it is easy to see that each operator $\Omega_{k i_3 i_4 i_5 \cdots}$ is an unambiguous unitary map $U$
with probability $p \left( k, i_3, i_4, \cdots \right)$ that is given by
\begin{equation}
p \left( k, i_3, i_4, \cdots \right) = |\omega_{k i_3 i_4 i_5 \cdots}|^2  . \label{pkiii}
\end{equation}
Since $q= \sum_k p_k$, using the Eqs. from (\ref{omegakthetak}) to (\ref{pkiii}), we have
\begin{equation}
q = \sum_{k,i_3,i_4,\cdots} p \left( k, i_3, i_4, \cdots \right) .
\end{equation}
The quantum operation $\mathcal{E}^{\prime}$ that is defined by its operation elements
$\{ \Omega_{k i_3 i_4 i_5 \cdots } \}$ is certainly different from $\mathcal{E}$, however both $\mathcal{E}^{\prime}$ and $\mathcal{E}$
are unambiguous unitary quantum channels representing the same unitary channel $\mathcal{U}$ with the same probability $q$.

The advantage to use $\mathcal{E}^{\prime}$ is that its operation elements
have the simple form given in (\ref{omegakiii}), i.e.,
each operation element $ \Omega_{k i_3 i_4 i_5 \cdots }$ is
a direct product of the unitary map $U$ and bras and kets in the appropriate Hilbert spaces.
Usually each system is held by a different observer, in order to accomplish the UUQC $\mathcal{E}$, some shared
resources (including entangled quantum states, quantum channels, classical channels and classical randomness) are needed.
Does the refined operation $\mathcal{E}^{\prime}$ that represents the same unitary channel $\mathcal{U}$ with the same probability
as $\mathcal{E}$ require more resources to accomplish? The answer is NO.
The only difference between $\mathcal{E}^{\prime}$ and $\mathcal{E}$ is
that $\mathcal{E}^{\prime}$ contains additional local measurements (see (\ref{omegarefine})),
but the results of local measurements
need not be communicated, therefore the shared resources needed are the same.
Furthermore, it is obvious from (\ref{omegakiii}) that for fixed $i_3,i_4,i_5,\cdots$,
the operators $ \Omega_{k i_3 i_4 i_5 \cdots }$ with different
values of $k$ are proportional to each other; therefore we can
combine them by defining the operation
$\mathcal{E}^{\prime}$ in terms of
$\{ \Omega_{i_3 i_4 i_5 \cdots} \}$ as its operation elements instead,
\begin{equation}
\Omega_{i_3 i_4 i_5 \cdots} =  \sqrt{\sum_k  |\omega_{k i_3 i_4 i_5 \cdots}|^2}
U \otimes \left( \left| i_4 \right> \otimes \left| i_6 \right> \otimes \cdots \right)
\otimes \left( \left< i_3 \right| \otimes \left< i_5 \right| \otimes \cdots \right) .
 \label{omegakiii22}
\end{equation}
We formally state the above result as a proposition.

{\bf Proposition 4.}
If an UUQC $\mathcal{E}_0$, which is a probabilistic $\mathcal{U}$ from $\hat{\mathcal{H}}_1^s$ to $\hat{\mathcal{H}}_2^s$
with probability $q$ ($\mathcal{E}_0= \mathcal{E}_{0\left(U, q \right)}$), can be accomplished with certain shared resources,
then there exists a quantum operation $\mathcal{E}$ that has the following three properties:
(a) it is also a probabilistic $\mathcal{U}$ with the same probability $q$ ($\mathcal{E}= \mathcal{E}_{\left(U, q \right)}$),
(b) $\mathcal{E}$ can also be accomplished with the same amount of shared resources,
and (c) the operation elements $\Omega_k$ of $\mathcal{E}$ can be chosen to satisfy
\begin{equation}
\Omega_k = U \otimes \Theta_k
\end{equation}
with $U$ the unitary map from $\mathcal{H}_1^s$ to $\mathcal{H}_2^s$, and each $\Theta_k$ a rank-one operator
(a product of bras and kets in the appropriate Hilbert spaces),
\begin{equation}
\Theta_k = \omega_k \left( \left| i_4 \right> \otimes \left| i_6 \right> \otimes \cdots \right) \nonumber
\otimes \left( \left< i_3 \right| \otimes \left< i_5 \right| \otimes \cdots \right)
\end{equation}
where $\omega_k$ is a nonzero complex number with $k$ the combination of $i_3,i_4,i_5,\cdots$,
and $\{ \left| i_l \right>\}$ is an arbitrary basis of $\mathcal{H}_l$ ($l=3,4,5,\cdots$) we choose.
When we deal with an optimization problem, this proposition ensures a simple form for the operation elements of an optimum protocol
and simplifies certain measurements involved, as we shall see later.

Suppose $\mathcal{E}$ is an unambiguous unitary quantum channel from $\hat{\mathcal{H}}_1^s$ to $\hat{\mathcal{H}}_2^s$,
$\mathcal{E} =\mathcal{E}_{\left( U, q \right)}$.
Let $\mathcal{H}_a$ be a Hilbert space distinct from any of those we have been considering. We can define a linear superoperator
$\mathcal{I}_a \circ \mathcal{E} $ that maps operators on $\mathcal{H}_a \otimes \mathcal{H}_1^s$
to operators on $\mathcal{H}_a \otimes \mathcal{H}_2^s$ according to
\begin{equation}
\mathcal{I}_a \circ \mathcal{E} \left( \sigma_a \otimes \sigma_1 \right) = \sigma_a \otimes \mathcal{E} \left( \sigma_1 \right)
\end{equation}
where $\sigma_a$ ($\sigma_1$) is any operator on $\mathcal{H}_a$ ($\mathcal{H}_1^s$); i.e., $\mathcal{I}_a \circ \mathcal{E} $ is
the tensor product of an identity map $\mathcal{I}_a$ (on $\hat{\mathcal{H}}_a$) with $\mathcal{E}$. From proposition 3 and the definition of the superoperator
$\mathcal{I}_a \circ \mathcal{E} $ on combined system, we can prove the following proposition.

{\bf Proposition 5.}
The tensor product $\mathcal{I}_a \circ \mathcal{E} $ of an identity map $\mathcal{I}_a$ (on $\hat{\mathcal{H}}_a$) with an UUQC $\mathcal{E}$ ($ =\mathcal{E}_{\left( U, q \right)}$)
from $\hat{\mathcal{H}}_1^s$ to $\hat{\mathcal{H}}_2^s$ is an UUQC
from $\hat{\mathcal{H}}_a \otimes \hat{\mathcal{H}}_1^s$ to $\hat{\mathcal{H}}_a \otimes \hat{\mathcal{H}}_2^s$
with the same probability $q$.

\subsection{Equivalence between unambiguous unitary quantum channels and uniformly entangled states}  ~\label{UUQCandues}
\nopagebreak

A bipartite pure state in a Hilbert space $\mathcal{H}_{12}$ of dimension $d_1 d_2$ is called a uniformly
entangled state (UES) of (Schmidt) rank $d$ ($d\leq d_1, d\leq d_2$) if it has $d$ nonzero Schmidt coefficients
that are all equal to $\frac{1}{\sqrt{d}}$.
When $d=d_1=d_2$, a UES is just a maximally entangled state (see, e.g. \cite{werner01,wgc03,szk02}).
Any UES of rank $d$ is related by a local unitary operation to the state
\begin{equation}
\left| \Phi_d \right> \equiv \frac{1}{\sqrt{d}} \sum_{i=1}^{d} \left| i \right>_1 \left| i \right>_2
\end{equation}
where only the first $d$ basis states of each system appear in the summation.

There is an equivalence between an unambiguous unitary quantum channel and a uniformly entangled state.
If Alice and Bob have an unambiguous unitary quantum channel $\mathcal{E}_{\left(U, q \right)}$ from $\hat{\mathcal{H}}_1^s$ to $\hat{\mathcal{H}}_2^s$,
with $\mathcal{H}_1^s$ a $d$-dimensional subspace of the Hilbert space of system 1 held by
Alice and  $\mathcal{H}_2^s$ a $d$-dimensional subspace of the Hilbert space of system 2 held by Bob,
then they can use it to create a uniformly entangled state of rank $d$ with the same probability $q$ by the following steps:
Alice first prepares two particles (in her hand) in the state $\left| \Phi_d \right>$, then she sends one particle through the UUQC.
As a result of proposition 5 we know they will obtain the UES of rank $d$ shared between them with probability $q$.
On the other hand, if they can have $\left| \Phi_d \right>$ with some probability $q$,
then supplemented with $2 \log_2 d$ cbits, they can establish an unambiguous
unitary quantum channel with the same probability by a usual teleportation scheme. The equivalence is explicitly shown as follows.

\bigskip

\parbox[]{3cm}{\sloppy An UUQC, $\mathcal{E}_{\left( U, q, d \right)}$}
\hfill
$\Longrightarrow$
\hfill
\parbox[]{3cm}{\sloppy a UES $\left| \Phi_d \right>$ with probability $q$}

\nopagebreak
\parbox[]{3cm}{\sloppy An UUQC, $\mathcal{E}_{\left( U, q, d \right)}$}
\hfill
$\stackrel{\textrm{cbits}}{\Longleftarrow}$
\hfill
\parbox[]{3cm}{\sloppy a UES $\left| \Phi_d \right>$ with probability $q$}


\nopagebreak
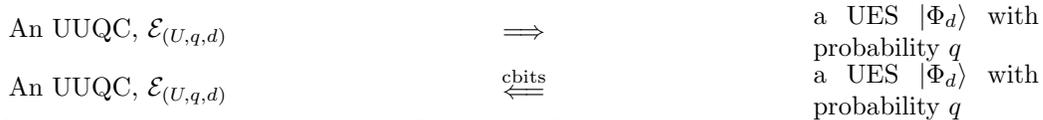
\noindent{FIG. 6.  Equivalence between an UUQC and a UES}

\bigskip

A consequence of the equivalence is the following proposition.

{\bf Proposition 6.}
Suppose Alice and Bob have some shared quantum resource (which could be a quantum channel or an entangled state or both),
they can turn their resource into an unambiguous unitary quantum channel
$\mathcal{E}_{\left(U, q, d \right)}$ by means of local operations and classical communication (LOCC) if and only if
they can obtain from the same shared quantum resource a uniformly entangled state $\left| \Phi_d \right>$ of rank $d$
with the same probability $q$ by means of LOCC.

\subsection{Examples of unambiguous unitary quantum channels}  ~\label{exampleUUQC}
\nopagebreak

We have discussed unambiguous unitary quantum channels in a very general context without referring to the observers who hold the systems
or the shared resources that are used to establish the unambiguous quantum channels.  When we specify the particular resources shared
and the particular relations between the systems and the observers, we obtain particular examples of unambiguous unitary quantum channels.

A perfect quantum channel from Alice to Bob, which can transfer a particle from Alice to Bob perfectly, is an example of an unambiguous
unitary channel (with unity probability).
\begin{center}
\setlength{\unitlength}{0.05in}
\begin{picture}(15,6)(0,-5)
\thicklines

\put(0,-3){\vector(1,0){10}}
\put(10,-3){\line(1,0){5}}
\put(0,-1){\makebox(0,0)[]{Alice}}
\put(0,-4.5){\makebox(0,0)[]{1}}
\put(15,-1){\makebox(0,0)[]{Bob}}
\put(15,-4.5){\makebox(0,0)[]{2}}

\label{pictpqc0009}
\end{picture}

\nopagebreak
\noindent{FIG. 7.  A perfect quantum channel}
\end{center}

In general, it is not necessary to transfer a particle in order to construct an UUQC. Suppose Alice and Bob share a uniformly entangled state of Schmidt
rank $d$, together with $2 \log_2 d$ bits of classical communication, Alice can teleport any state in a $d$-dimensional Hilbert space to Bob's particle.
Quantum teleportation with uniformly entangled state as the quantum resource is certainly an UUQC with probability $1$.

When Alice and Bob share a partially entangled state as well as a classical channel,
and Alice still wants to send Bob an unknown state, they
want the state to be sent with fidelity $1$ when their protocol
succeeds. This is the case of unambiguous teleportation, which
is an UUQC with a probability less than $1$.

Suppose Alice and Bob share a noisy quantum channel that could cause error, Alice still wants to send Bob an arbitrary state in a particular Hilbert space
with fidelity $1$. The strategy based on error correction is as follows: Alice first encodes her message in a particular subspace of an enlarged system,
she sends the enlarged system through the noisy quantum channel, Bob performs error correction to correct the errors and decode the message. The combined
operation of the noisy quantum channel and the error correction can also be considered as an unambiguous unitary quantum channel with probability $1$.
Unambiguous error correction with probability less than $1$, which is an UUQC, will be discussed later.

When Alice and Bob share a non-maximally entangled state, they cannot perform faithful dense coding with probability $1$ in general, however they can perform
unambiguous dense coding.  It will be shown that unambiguous dense coding with certain requirement can also be viewed as an UUQC.

\section{Unambiguous teleportation}  ~\label{contele}
\nopagebreak

Unambiguous teleportation
\cite{horodeckimpr, bhm, griffiths2005, mh, slkp, band, ap, pa, rdg, kkj, kkaj, uhlmann, lsl, rdf, gour} with any shared
entangled state is an example of an UUQC with the observers and
quantum resources particularly specified. Now suppose Alice has
particle 1 (system $\mathcal{S}_1$) and particle 3 (system
$\mathcal{S}_3$), Bob has particle 2 (system $\mathcal{S}_2$). We
will denote by $\mathcal{H}_1$ the $d$-dimensional Hilbert space of
particle 1, and by $\mathcal{H}_2$ ($\mathcal{H}_3$) the
Hilbert space of particle 2 (3).   Particle 1 is in
an unknown state $\left| \psi_1 \right> \in \mathcal{H}_1$.
Particles 2 and 3 are in a shared entangled state $\rho_{23}$, which
is the quantum resource shared between Alice and Bob. With this
shared entangled state and additional classical
communication, Alice wants to send Bob the unknown state $\left|
\psi_1 \right>$ with unity fidelity. They need to find the maximal probability of success and the protocol
to achieve it.

Their task is to establish an UUQC using the shared entangled state and enough additional classical communication.
From proposition 6, we know that such an UUQC can be established if and only if a UES of rank $d$ can be obtained with the same probability from
$\rho_{23}$ by LOCC. Therefore the problem of unambiguous teleportation is equivalent to the problem of entanglement purification
with perfect fidelity.

Suppose the shared state is a general mixed state $\rho_{32}$,
which can be viewed as the reduced density matrix of a tripartite pure state $\left| \Psi \right> _{325} $.
\begin{center}
\setlength{\unitlength}{0.05in}
\begin{picture}(36,18)(-2,-18)
\thicklines
\put(5,-2){\vector(-1,0){5}}
\put(2,-0.5){\makebox(0,0)[]{$4$}}
\put(5,-4){\framebox(4,4)[]{$A_x$}}
\put(19,-1){\vector(-1,0){5}}
\put(14,-1){\line(-1,0){5}}
\put(15,0.5){\makebox(0,0)[]{$1$}}
\put(19,-3){\vector(-1,0){5}}
\put(14,-3){\line(-1,0){5}}
\put(15,-4.5){\makebox(0,0)[]{$3$}}
\put(19,-3){\line(0,-1){5}}
\put(17,-12){\framebox(4,4)[]{$\left| \Psi \right>$}}
\put(19,-12){\line(0,-1){5}}
\put(19,-17){\vector(-1,0){5}}
\put(15,-15.5){\makebox(0,0)[]{$2$}}
\put(14,-17){\line(-1,0){5}}
\put(5,-19){\framebox(4,4)[]{$B_x$}}
\put(5,-17){\vector(-1,0){5}}
\put(17,-10){\vector(-1,0){17}}
\put(10,-8.5){\makebox(0,0)[]{$5$}}

\label{pict51}
\end{picture}

\nopagebreak
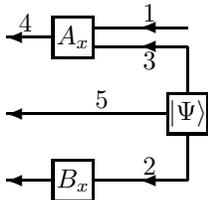
\noindent{FIG. 8.  Alice and Bob share a mixed state. }
\end{center}
Suppose $\{A_x \otimes B_x, \; | x=1,\cdots,n \}$ is the set of operation elements for the LOCC operation $\emph{L}$ that realizes the unambiguous
teleportation of an unknown state of particle 1.
From proposition 4, leg $4$ is detached from legs $1$ and $3$, hence the operators $A_x$ can be refined to rank-one operators
without reducing the probability of success for unambiguous teleportation.

{\bf Proposition 7: }
When the shared entangled state is generally a mixed state,
the measurement performed by Alice in
unambiguous teleportation has the following
property: the POVM elements $G_x$ and the corresponding Kraus operators $A_x$
associated with the success cases (i.e., $x \geq 1$) can be chosen as
rank-one operators without reducing the probability of success.

It is a nontrivial observation that the POVM elements ($G_x =A_{x}^{\dag} A_{x}$)
involved in the realization of an unambiguous teleportation (as well as in the distillation protocol
to obtain a UES)
can be chosen to be rank-one without reducing the probability of success.

It should be pointed out that in previous references the POVM elements
$G_x$ ($x \geq 1$) are usually assumed to be rank-one operators without a detailed proof,
here we provide a rigorous proof based on the property of a probabilistic unitary map.

Since $A_{x}$ is a rank-one operator, it can be written as
$A_{x}= \left|\upsilon_x\right>_4 \otimes \left< a_{x} \right|$, where $\left< a_{x} \right|$ is
a bra in $\mathcal{H}^{\dagger}_{13}$. Let $P_3^{x}$ denote the
projector onto the $d-$dimensional subspace $\mathcal{H}_3^s$ of system 3
spanned by the $d$ Schmidt basis states (that correspond to nonzero Schmidt coefficients) of $\left| a_x \right>$,
and let $P_2^x$ denote the
projector onto the $d-$dimensional support space of $B_x$.
From proposition 4 we know that leg 5 is detached from the whole diagram, which means that
$(P_3^{x} \otimes P_2^x) \rho_{32} ( P_3^{x} \otimes P_2^x)$ is a pure state. Hence we have the following result.

{\bf Proposition 8:}
The probability for unambiguous teleportation of a state in a $d$-dimensional subspace
is nonzero if and only if
there exists a $d$-dimensional subspace for each system of the shared entangled state,
such that we have a nonzero probability to
get a pure state of Schmidt rank $d$ by projecting the shared state
onto the $d \times d$ dimensional subspace of the two systems.

A similar result about entanglement purification was also obtained in (\cite{horodeckimpr}).

\section{Quantum error correction}  ~\label{qec}
\nopagebreak

The combined operation of quantum error correction \cite{bdsw, cs, kl, crss, gottesman96, steane9601, shor96} and the noisy
channel (that causes the error) can be viewed as an unambiguous
unitary quantum channel with probability $1$. All the measurements
involved in the combined operation
can be removed, since (1) the errors can be introduced by an
interaction with an environment that is initially in a fixed state, and (2) the
error correction operation can also be implemented by a unitary
interaction with an ancilla system initially in a given pure state.

\begin{center}
\setlength{\unitlength}{0.05in}
\begin{picture}(37,13)(2,-9)
\thicklines
\put(2,-5){\vector(1,0){4}}
\put(6,-5){\line(1,0){2}}
\put(8,-7){\framebox(4,4)[]{$C$}}
\put(12,-5){\vector(1,0){4}}
\put(16,-5){\line(1,0){2}}
\put(18,-7){\framebox(4,4)[]{$E$}}
\put(20,-3){\vector(0,1){3}}
\put(22,-5){\vector(1,0){4}}
\put(26,-5){\line(1,0){2}}
\put(28,-7){\framebox(4,4)[]{$R$}}
\put(30,-3){\vector(0,1){3}}
\put(32,-5){\vector(1,0){5}}
\put(39.5,-5){\makebox(0,0)[]{$\left| \psi \right>$}}
\put(34,-3.5){\makebox(0,0)[]{\small $2$}}
\put(6,-3.5){\makebox(0,0)[]{\small $1$}}
\put(24,-3.5){\makebox(0,0)[]{\small $2$}}
\put(14,-3.5){\makebox(0,0)[]{\small $2$}}
\put(0,-5){\makebox(0,0)[]{$\left| \psi \right>$}}
\put(30,1.5){\makebox(0,0)[]{$r$}}
\put(20,1.5){\makebox(0,0)[]{$e$}}
\label{pictUUQCerrcorr}
\end{picture}

\nopagebreak
\noindent{FIG. 9.  Quantum error correction. }
\end{center}

In the above diagram, $C$ denotes the encoding operation, an isometry that maps a $d$-dimensional Hilbert space $\mathcal{H}_1$
onto the code space $\mathcal{H}_2^c$ that is a subspace of the Hilbert space $\mathcal{H}_2$ of
the principal system (which consists of all the particles used to encoding the message).
$E$ denotes an isometry that maps the Hilbert space $\mathcal{H}_2$ into the product $\mathcal{H}_2\otimes \mathcal{H}_e$ of the Hilbert space
$\mathcal{H}_2$ and the Hilbert space $\mathcal{H}_e$ of the environment $e$ whose interaction with the principal system causes the error. It is equivalent to
describe the noise by a quantum operation $\mathcal{E}$, for any state $\rho_2$ in $\hat{\mathcal{H}}_2$ we have
$\mathcal{E} \left( \rho_2 \right) =\sum_i E_i \rho_2 E_i^{\dag}$ where $\{ E_i \}$ are error operation elements which is related to the isometry $E$ by
$E_i \equiv \left< e_i \right| E$ with $\{ \left| e_i \right> \}$ a set of basis states of the environment.

The error-correction operation is usually described as a two stage process, the syndrome measurement on the principal system is described by
measurement operators $\{ M_j \}$, and the error-correction step is described by a set of corresponding conditional unitary operations $U_j$.
However this whole error-correction operation can be described by a unitary operation on the principal system plus an
ancilla system $r$ which is initially in
a fixed state. The error-correction operation is equivalently described by an isometry $R$
that maps the Hilbert space $\mathcal{H}_2$ into the product space $\mathcal{H}_{2r}$ of the Hilbert space $\mathcal{H}_2$ and the Hilbert
space $\mathcal{H}_r$ of the ancilla system. The projector onto the code space $\mathcal{H}_2^c$ is denoted by $P_2^c$.

The operator $RE$ can be viewed as a (unambiguous) unitary quantum map with probability $1$ from the code space $H_2^c$ to itself,
we also have $P_2^c R E P_2^c = R E P_2^c$ since the probability of successful error-correction is $1$.  From proposition $2$ we have
\begin{equation}
R E P_2^c  = \left| \Theta_{er} \right> \otimes P_2^c  \label{UUQCerrcor1}
\end{equation}
where $\left| \Theta_{er} \right>$ is a normalized ket in the product space $\mathcal{H}_{e} \otimes \mathcal{H}_{r}$.
Let ${\left| e_i \right>}$ be the set of basis states of $\mathcal{H}_e$ corresponding to the set of linearly independent error operation elements
$\{ E_i \}$, namely $E_i = \left< e_i \right| E$.  Using the fact that the product $R^{\dag}R$ is an identity operator
on $\mathcal{H}_2$ (since $R$ is an isometry), from (\ref{UUQCerrcor1}) we immediately obtain the well-known condition
\begin{equation}
P_2^c E_j^{\dag} E_i P_2^c = h_{ji} P_2^c
\end{equation}
with $h_{ji} \equiv \left<e_i \right| \rho_e \left| e_j \right>$, where $\rho_e \equiv tr_r \left( \left| \Theta_{er} \right> \left< \Theta_{er} \right| \right)$
is a density operator of the environment $e$.
If $\{ \left|e_i \right> \}$ is the basis in which the bipartite pure state $\left| \Theta _{er} \right>$ has its Schmidt form,
then the matrix $h$ is diagonal.
We have proved the well known condition for the correctable errors by viewing the combined process of error correction and noisy channel
as an unambiguous unitary quantum channel with probability $1$. Our approach here is
different from the entropic approach developed in \cite{ncsb97,sw01,sw02}.

The operation $\mathcal{E}$ that describes the noise has been assumed to be trace-preserving,
if it is trace-decreasing, all results hold except that $\left| \Theta _{er}\right>$ and $\rho_e$ are not normalized.

It is convenient to consider quantum error correction in the following scenario. Alice and Bob share a noisy quantum channel $\mathcal{E}$,
Alice first encodes her message state into a subspace of the Hilbert space of the principal system
(this step is represented by the isometry $C$), then she sends the
principal system to Bob through the noisy quantum channel
(this step is represented by the isometry $E$ in the diagram).
After Bob receives the principal system, he performs error correction operation $\mathcal{R}$
(this step is represented by the isometry $R$ in the diagram).

Sometimes the error cannot be corrected with certainty, but only with a probability $p\leq 1$, however we need to know
whether the error is corrected successfully. We refer to this as unambiguous error correction.
The combined operation of the noisy quantum channel and the error correction
can still be viewed as an unambiguous unitary channel with some probability $p$.
Suppose $Dim \left( \mathcal{H}_1 \right)= Dim \left( \mathcal{H}_2^c \right)=d$, Alice can prepare a uniformly entangled state
$\left| \Phi_d \right>$ of two particles $a$, $b$, and feed only particle $b$ through the encoding operation
(corresponding to the isometry $C$) and noisy channel $\mathcal{E}$,
the resulting state is
$I_a \circ \mathcal{E}( I_a \otimes C \left| \Phi_d \right> \left< \Phi_d \right| I_a \otimes C^{\dag})$.
From proposition 6, we know that
the probability of obtaining $\left| \Phi_d \right> $ from the resulting state
with only local operations on particle $b$ (no classical information needs to be communicated in this case)
is equal to the probability $p$ that the error can be corrected.
We have the following proposition.

{\bf Proposition 9.}
The error caused by a noisy channel $\mathcal{E}$ can be corrected with probability $p$ ($p\leq 1$) if and only if
the uniformly entangled state $\left| \Phi_d \right> $ can be obtained from the bipartite state
$I_a \circ \mathcal{E} ( I_a \otimes C \left| \Phi_d \right> \left< \Phi_d \right| I_a \otimes C^{\dag})$ ($\in \mathcal{H}_{ab}$)
with probability $p$ by local operations on particle $b$.

Since local operations cannot increase entanglement, we immediately get the following corollary.

{\bf Corollary 10:}
The error caused by a noisy channel $\mathcal{E}$ can be corrected with certainty only if the bipartite state
$I_a \circ \mathcal{E} ( I_a \otimes C \left| \Phi_d \right> \left< \Phi_d \right| I_a \otimes C^{\dag})$ ($\in \mathcal{H}_{ab}$)
is a uniformly entangled state of rank $d$.

\section{Unambiguous dense coding} ~\label{cdc}
\nopagebreak

Dense coding provides a method whereby a shared, entangled resource may be used to increase the classical capacity of a quantum channel. Since its discovery by Bennett and Wiesner \cite{bw}, a number of generalizations have been discussed in the literature. When the shared entangled state of Schmidt rank $D$ is not
maximally entangled, many interesting results have been obtained \cite{be,hjsww,mro, pati}.

The problem of unambiguous dense coding in its general form is as follows: Alice and Bob share an entangled state, and wish to use it
to communicate one of various possible classical messages from Alice to Bob. To do so, Alice performs an arbitrary quantum operation on
her half of the entangled state, conditioned on the message she has chosen to send, and then sends her half to Bob through a perfect
quantum channel. Bob then performs a measurement on the full system to attempt to ascertain which message Alice chose to send.
Alice's set of operations and Bob's measurement may in principle be chosen from the most general class of possibilities. Given these choices,
their probability of success may range from zero to one, and may depend on which message was actually sent. Bob, however, must know
with certainty whether or not he has successfully determined Alice's message. This is what the term \lq \lq unambiguous" means.

Unambiguous dense coding is closely related to
state discrimination \cite{chefles,ivanovic, dieks, peres, jaeger, duan, chefles2, sun1}.
A more general discussion about unambiguous dense coding is given in \cite{wcsg},
while in this paper we only consider the case when unambiguous dense coding can be considered as an unambiguous unitary quantum
channel, namely when all Alice's messages are required to be transmitted
with equal success probability.

Suppose Alice and Bob share an entangled state of Schmidt rank $D$, $\left|E_D \right> = \sum_{i=1}^{D} \lambda_i \left| i \right> \left| i \right>$
($\lambda_1 \geq \lambda_2 \geq \cdots \geq \lambda_D >0$).
Alice wants to send $2 \log_2 D$ classical bits of information to Bob. Since the shared state is not a maximally entangled state, the probability of
success is less than one.  We only consider the case when all $D^2$ classical messages
should be sent through with equal probability $p$.
We like to get the maximal success probability.

This problem can be nicely formulated as an unambiguous unitary quantum channel.
\begin{center}
\setlength{\unitlength}{0.045in}
\begin{picture}(47,19)(0,-18)
\thicklines
\put(4,-4){\vector(1,0){3}}
\put(7,-4){\line(1,0){2}}
\put(9,-7){\framebox(4,6)[]{$A$}}
\put(13,-2){\line(1,0){4}}
\put(21,-2){\vector(-1,0){4}}
\put(21,-4){\framebox(4,4)[]{$E_D$}}
\put(25,-2){\vector(1,0){4}}
\put(29,-2){\line(1,0){4}}
\put(13,-6){\vector(1,0){10}}
\put(23,-6){\line(1,0){10}}
\put(33,-7){\framebox(4,6)[]{$B$}}
\put(37,-4){\vector(1,0){4}}
\put(45,-4){\makebox(0,0)[]{\small $r \left|x \right>$}}
\put(2,-4){\makebox(0,0)[]{\small $ \left|x \right>$}}

\put(4,-14){\vector(1,0){3}}
\put(7,-14){\line(1,0){2}}
\put(9,-17){\framebox(4,6)[]{$\tilde{A}$}}
\put(13,-12){\vector(1,0){4}}
\put(18,-10.5){\makebox(0,0)[]{\small $1$}}
\put(21,-12){\line(-1,0){4}}
\put(21,-14){\framebox(4,4)[]{$\tilde{E}_D$}}
\put(25,-12){\vector(1,0){4}}
\put(29,-12){\line(1,0){4}}
\put(13,-16){\vector(1,0){10}}
\put(18,-17.5){\makebox(0,0)[]{\small $2$}}
\put(23,-16){\line(1,0){10}}
\put(33,-17){\framebox(4,6)[]{$B$}}
\put(37,-14){\vector(1,0){4}}
\put(45,-14){\makebox(0,0)[]{\small $r \left|x \right>$}}
\put(2,-14){\makebox(0,0)[]{\small $ \left|x \right>$}}
\label{pictUUQCdensecoding}
\end{picture}

\nopagebreak
\noindent{FIG. 10.  Unambiguous dense coding. }
\end{center}
In FIG. 10, $E_D$ denotes the ket $\left| E_D \right>$, $\tilde{E}_D$ denotes the diagonal matrix that is given from $E_D$ by a
transpose with respect to the basis of Alice's particle, $\tilde{E}_D = diag \{ \lambda_1, \cdots, \lambda_D \}$. When the message to be
sent is $x$ $\{x=1,\cdot,D^2 \}$ , then Alice performs a corresponding operation $A_x$ on her particle of the shared state.
The operation $A_x$ is not necessary to be a unitary operation, it is generally a Kraus operator of a POVM measurement, the only requirement on
$A_x$ is that it represents a trace-non-increasing operation, namely, $A_x^{\dag} A_x  \leq I$. The operator $A$ in the diagram is defined as
$A \equiv \sum_{x=1}^{D^2} A_x \otimes \left< x \right|$, with $\{ \left|x \right> \}$ being a basis in the product space
$\mathcal{H}_1 \otimes \mathcal{H}_2$ where $\mathcal{H}_2$ is an exact copy of the Hilbert space $\mathcal{H}_1$.
Each operator $A_x$ on $\mathcal{H}_1$ can be transformed into a ket $\tilde{A}_x$ in
$\mathcal{H}_1 \otimes \mathcal{H}_2$  by transposing only the bras. The object defined by
$\tilde{A} \equiv \sum_{x=1}^{D^2}  \tilde{A}_x \otimes \left< x \right|$
can be viewed as an operator acting on the vectors in a $D^2$-dimensional space $\mathcal{H}_1 \otimes \mathcal{H}_2$,
i.e., $\tilde{A} \in \hat{\mathcal{H}}_1 \otimes \hat{\mathcal{H}}_2$.
$B$ is a Kraus operator of some POVM measurement, $B^{\dag} B \leq I_{12}$, where $I_{12}$ ($I_1$, $I_2$) is the identity operator on
$\mathcal{H}_{12}$ ($\mathcal{H}_{1}$, $\mathcal{H}_{2}$).
The upper diagram and the lower diagram are equal as the consequence of a property of the atemporal diagrams, namely an atemporal diagram remains
unchanged if the direction of an inner line is reversed.

The lower (or upper) diagram in FIG. 10 can be viewed as an unambiguous unitary quantum map from
$\mathcal{H}_1 \otimes \mathcal{H}_2$ to itself
with probability $p=|r|^2$. So we have
$B \left( \tilde{E}_D \otimes I_2 \right) \tilde{A} = r I_{12}$ which gives
\begin{equation}
B = r \tilde{A}^{-1} \left( \tilde{E}_D^{-1} \otimes I_2 \right)  \; . \label{UUQCdensecoding1}
\end{equation}
The condition $B^{\dag} B \leq I_{12}$ and (\ref{UUQCdensecoding1}) implies
\begin{equation}
\frac{1}{|r|^2} \left\{ \tilde{A}^{\dag} \left( \tilde{E}_D \otimes I_2 \right) \left( \tilde{E}_D \otimes I_2 \right) \tilde{A}  \right\} \geq I_{12} \; ,
\label{UUQCdens255}
\end{equation}
which in turn implies \cite{equeigen}
\begin{equation}
\frac{1}{|r|^2} \left\{ \left( \tilde{E}_D \otimes I_2 \right) \tilde{A} \tilde{A}^{\dag} \left( \tilde{E}_D \otimes I_2 \right)  \right\} \geq I_{12} \; .
\label{UUQCdensecoding3}
\end{equation}
The conditions $A_x^{\dag} A_x  \leq I_1$ is equivalent to $\left( A_x^{\dag} A_x \right)^* \leq I_1$ that implies
\begin{equation}
tr_2 \left( \tilde{A} \tilde{A}^{\dag} \right)  \leq D^2 I_1  \; .  \label{UUQCdensecoding5}
\end{equation}
Tracing (\ref{UUQCdensecoding3}) over $\mathcal{H}_2$, together with (\ref{UUQCdensecoding5}) we have
\begin{equation}
|r|^2 I_1 \leq D \tilde{E}_D^2
\end{equation}
which implies $p=|r|^2 \leq D \lambda_D^2$.  The maximal probability of success is $D \lambda_D^2$, since it is achievable by an optimal
protocol given as follows.  Alice chooses $A_x$ to be orthogonal unitary operators
(i.e., $A_x^{\dag} A_x = I_1$ and $Tr \{ A_x^{\dag} A_y \}= \delta_{xy} D$), and Bob's operation $B$ is performed in two steps.
Bob first tries to get a maximally entangled state by performing a one-shot distillation, which is a measurement on Bob's particle
with the Kraus operators given by
$K=\sum_{i=1}^{D} \frac{\lambda_D}{\lambda_i} \left|i\right> \left< i \right|$ (success), and
$\sqrt{1-K^{\dag}K}$ (failure). Bob's first step can
be done before or after he receives Alice's particle, his first step will succeed with probability $D \lambda_D^2$. After he receives
Alice's particle, Bob performs a projective measurement trying to distinguish the $D^2$ states
$\left( K\otimes A_x \right) \left| E_D \right>$ ($x=1,\cdots,D^2$)
which are orthogonal to each other, this step will succeed with certainty. Thus the
total probability of success is $D\lambda_D^2$, and this protocol is optimal.
We have proved the following proposition.

{\bf Proposition 11: }
The maximal success probability is equal to $D \lambda_D^2$, and it can be achieved by
a protocol that involves a local entanglement distillation procedure performed by Bob and a follow-up standard
dense coding protocol.

Although we require all messages should be transmitted with the same probability, we don't impose restrictions
(like unitarity) on Alice and Bob's local operations.

\section{Conclusion}
\nopagebreak

\subsection{Summary}
\nopagebreak

Unambiguous unitary maps (UUMs) and unambiguous unitary quantum channels (UUQCs) are introduced.
Some properties of UUMs and UUQCs are derived. The Kraus operators
of an UUQC which contribute to the success probability are UUMs.
An UUM as well as an UUQC has no preference on the input state, and it has certain simple form
when conditional on the subspaces we are interested in.
For any UUQC, there exists
another UUQC, which represents the same unitary channel with the
same probability and can be achieved with the same amount of shared
resources, and is much simpler as it can be written as a product of
the unitary map and a rank-one operator. Connection between
an UUQC and a uniformly entangled state is also obtained.

Unambiguous teleportation, quantum error corrections and unambiguous dense coding
are discussed as examples of UUQCs.
A rigorous proof is provided for the fact that
the POVM elements and the corresponding Kraus operators that contribute to the
success probability in unambiguous teleportation can always be chosen as rank-one operators.
For the first time a necessary and sufficient condition
for a set of errors to be correctable with a nonzero
probability $p \leq 1$ is obtained, and a necessary
condition as a corollary is also obtained.
Dense coding with a non-maximally entangled state, when all the classical messages
are required to have equal success probability of transmission, is discussed;
the maximal success probability is derived and the optimum protocol is also given.

\subsection{Open questions}
\nopagebreak

The main examples of unambiguous unitary quantum channels discussed in this article
are very different in terms of the goal to achieve and the resources for use.
It would be interesting to ask whether we can find other quantum operations
as examples of unambiguous unitary quantum channels, and whether the formalism given
in this article is useful in finding novel applications in
quantum information processing.

The operations described by unambiguous unitary quantum channels are special in the sense that an unknown state should be
transmitted with perfect fidelity by these operations.
However, in the real world, we usually deal with operations that cannot preserve an unknown state with perfect fidelity.
It is interesting to ask whether we can have a similar set of results
if we consider a more general set of operations,
for example, those that transform an unknown state with a certain fidelity or with a fidelity no less than a certain value.

For unambiguous dense coding, we did not consider the more general case when each classical message can be transmitted with a
different probability of success. We are not sure whether the general case can be described similarly by
an unambiguous unitary quantum channel or even its generalized form.

\section*{Acknowledgments}
\nopagebreak

The research described here received support from
the National Fundamental Research Program (grant No. 2006CB921900),
the National Natural Science Foundation of China (grant No. 10604051), the
Chinese Academy of Sciences and the University of Science and
Technology of China.

\section*{Appendix: proof of propositions 1 and 2.}
\nopagebreak

Without assuming that $p$ is independent of the input, (\ref{pqcdef}) is rewritten as
\begin{equation}
Tr_{e_2} \left\{P_2^s \Omega \left| \psi \right\rangle \left\langle \psi \right| \Omega^{\dag} P_2^s \right\} = p (\psi) \; U \left| \psi \right\rangle \left\langle \psi \right| U^{\dag}
\label{apdx1}
\end{equation}
for any $\left| \psi \right\rangle \in \mathcal{H}^s_1$.  We choose a basis of $\mathcal{H}^s_1$ such that it includes $\left| \psi \right\rangle $,
the other $d-1$ orthonormal basis states are denoted as $\left| \psi^{\perp}_{\mu} \right\rangle$. Since $U$ is a unitary map from $\mathcal{H}^s_1$ to
$\mathcal{H}^s_2$, $\{U\left| \psi \right\rangle, \; U\left| \psi^{\perp}_{\mu} \right\rangle \;|\; \mu=1,\cdots,d-1 \}$ is a basis of $\mathcal{H}^s_2$
($d= Dim \left( \mathcal{H}^s_2 \right) =Dim \left( \mathcal{H}^s_2 \right)$). This basis can be used to expand the object
$P^s_2 \Omega \left| \psi \right\rangle$,
\begin{equation}
P^s_2 \Omega \left| \psi \right\rangle = U \left| \psi \right\rangle \otimes \Theta (\psi) + \sum_{\mu=1}^{d-1} U \left| \psi^{\perp}_{\mu} \right\rangle \otimes \Gamma_{\mu} (\psi) \; .
\label{thproof122}
\end{equation}
Substituting this expansion into (\ref{apdx1}) and comparing the terms, we have
\begin{eqnarray}
& p (\psi) = Tr_{e_2} \{ \Theta (\psi) \Theta^{\dag} (\psi) \} \label{apdx2}  \\
& Tr_{e_2} \{ \Gamma_{\mu} (\psi) \Gamma^{\dag}_{\mu} (\psi) \} =0 \; (\mu=1,\cdots,d-1). \label{apdx3}
\end{eqnarray}
From (\ref{apdx3}) we have $ \Gamma_{\mu} (\psi) =0$, therefore (\ref{thproof122}) gives
\begin{equation}
P^s_2 \Omega \left| \psi \right\rangle = U \left| \psi \right\rangle \otimes \Theta (\psi)  \; . \label{apdx4}
\end{equation}

Now we are going to show that $\Theta$ is independent of $\left| \psi \right>$.
$\{ \left| i \right\rangle \}$ is an arbitrary fixed basis of $\mathcal{H}^s_1$,
(\ref{apdx4}) gives
\begin{equation}
P_2^s \Omega \left| i \right\rangle = U \left| i \right\rangle \otimes \Theta_i \;. \label{apdx5}
\end{equation}
Using the expansion $\left| \psi \right\rangle =\sum_{i=1}^{d} \alpha_i \left| i \right\rangle$,
from (\ref{apdx4}) and (\ref{apdx5}), we obtain
\begin{equation}
\left( \sum_{i=1}^d \alpha_i U \left| i \right\rangle \right) \otimes \Theta(\psi) = \sum_{i=1}^d \alpha_i U \left| i \right\rangle \otimes \Theta_i
\end{equation}
which implies
\begin{equation}
\alpha_i \Theta(\psi) =\alpha_i \Theta_i \; .
\end{equation}
Choosing a particular $\left| \psi \right\rangle$ such that all $\alpha_i \neq 0$, we immediately have
$\Theta_i = \Theta$ for all $i=1,\cdots,d$. Therefore $\Theta(\psi)= \Theta$ and it is independent of the input state
$\left| \psi \right\rangle$. From (\ref{apdx2}) we know that $p$ is also independent of the input state.
Therefore proposition 1 is proved.

From (\ref{apdx4}) and the fact that $\Theta$ is independent of $\left| \psi \right>$,
we can choose a basis $\{ \left|i \right\rangle \}$ of $\mathcal{H}^s_1$ and obtain
\begin{equation}
P^s_2 \Omega \left| i \right\rangle \left\langle i \right| = (U  \left| i \right\rangle \left\langle i \right|) \otimes \Theta \; .
\end{equation}
Summation over $i$ gives (\ref{proposition1it3}).  The converse part of proposition 2 is obvious.


{\small
\begin{thebibliography}{99}

\bibitem{bbcjpw} C.H. Bennett, G. Brassard, C. Crepeau, R. Jozsa, A. Peres and W.K. Wootters, Phys. Rev. Lett. {\bf 70}, 1895 (1993).

\bibitem{horodeckimpr} M. Horodecki, P. Horodecki and R. Horodecki, Phys. Rev. A {\bf 60}, 1888 (1999);  quant-ph/9807091.

\bibitem{bhm} G. Brassard, P. Horodecki and T. Mor, IBM J.\ RES.\ \& DEV.\ {\bf 48} (1), 87 (2004).

\bibitem{shor95} P.W. Shor, Phys. Rev. A {\bf 52}, R2493 (1995).

\bibitem{steane96} A.M. Steane, Phys. Rev. Lett. {\bf 77}, 793 (1996).

\bibitem{griffiths2005} R.B. Griffiths, S. Wu, L. Yu and S.M. Cohen, Phys. Rev. A {\bf 73}, 052309 (2006);  quant-ph/0507215.

\bibitem{zb04} K. \.{Z}yczkowski and I. Bengtsson, Open Syst. Inf. Dyn., {\bf 11}, 3 (2004);  quant-ph/0401119.

\bibitem{griffiths2004} R.B. Griffiths, Phys. Rev. A {\bf 71}, 042337 (2005);  quant-ph/0409106.

\bibitem{NielsenChuang} Chapter 8 of M.A. Nielsen and I.L. Chuang,
\textit{Quantum Computation and Quantum Information} (Cambridge University Press, Cambridge, 2000).

\bibitem{jp} D. Jonathan and M.B. Plenio, Phys.\ Rev.\ Lett. {\bf 83}, 1455 (1999).

\bibitem{lopopescu} H.-K. Lo and S. Popescu, Phys.\ Rev.\ A {\bf 63}, 022301 (2001).

\bibitem{werner01} R.F. Werner, J. Phys. A: Math. Gen. {\bf 34}, 7081 (2001).

\bibitem{wgc03} A. W{\'o}jcik, A. Grudka and R.W. Chhajlany, Quant. Inf. Procc {\bf 2}, 201 (2003).

\bibitem{szk02} M. Sinolecka,  K. \.{Z}yczkowski and M. Ku{\'s},  Acta Phys. Pol. B {\bf 33}, 2081  (2002); quant-ph/0110082.

\bibitem{mh} T. Mor and P. Horodecki,  quant-ph/9906039.

\bibitem{slkp} W. Son, J. Lee, M.S. Kim and Y.-J. Park, Phys. Rev. A {\bf 64}, 064304 (2001).

\bibitem{band} S. Bandyopadhyay, Phys. Rev. A {\bf 62} 012308 (2000);  quant-ph/0002032.

\bibitem{ap} P. Agrawal and A.K. Pati, Phys. Lett. A {\bf 305} 1217 (2002); quant-ph/0210004.

\bibitem{pa} A.K. Pati and P. Agrawal, J. Opt. B, {\bf 6} S844 (2004).

\bibitem{rdg} L. Roa, A. Delgado and I. Fuentes-Guridi, Phys. Rev. A, {\bf 68}  022310 (2003).

\bibitem{kkj} Z. Kurucz, M. Koniorczyk and J. Janszky, Fortschr. Phys., {\bf 49} 1019 (2001);  quant-ph/0308020.

\bibitem{kkaj} Z. Kurucz, M. Koniorczyk, P. Adam and J. Janszky, J. Opt. B, {\bf 5} S627 (2003).

\bibitem{uhlmann} A. Uhlmann,  quant-ph/0301116; quant-ph/0407244.

\bibitem{lsl} C. Li, H.-S. Song and Y.-X. Luo, Phys. Lett. A {\bf 297} 121 (2002).

\bibitem{rdf} L. Roa, A. Delgado and I. Fuentes-Guridi, Phys. Rev. A {\bf 68}, 022310 (2003).

\bibitem{gour} G. Gour, Phys. Rev. A {\bf 70}, 042301 (2004).

\bibitem{bdsw} C.H. Bennett, D.P. DiVincenzo, J.A. Smolin and W.K. Wootters,
Phys. Rev. A {\bf 54}, 3824 (1996);  quant-ph/9604024.

\bibitem{cs} A.R. Calderbank and P.W. Shor, Phys. Rev. A {\bf 54}, 1098-1105 (1996);  quant-ph/9512032.

\bibitem{kl} E. Knill and R. Laflamme,
Phys. Rev. A {\bf 55}, 900 (1997);  quant-ph/9604034.


\bibitem{crss} A.R. Calderbank, E.M. Rains, P.W. Shor and N.J.A. Sloane,
IEEE Trans. Inform. Theory 44, 1369 (1998);  quant-ph/9605005.

\bibitem{gottesman96} D. Gottesman,
Phys. Rev. A 54, 1862 (1996); quant-ph/9604038.

\bibitem{steane9601} A.M. Steane,
Proc. Roy. Soc. London A 452 (1996), 2551-2577;
quant-ph/9601029.

\bibitem{shor96} P.W. Shor, 
Proc. 35th Ann. Symp. on Fundamentals of Computer Science (IEEE Press, Los Alamitos, 1996), pp. 56-65; quant-ph/9605011.

\bibitem{ncsb97} M.A. Nielsen, C.M. Caves, B. Schumacher and H. Barnum, quant-ph/9706064.

\bibitem{sw01} B. Schumacher and M.D. Westmoreland, quant-ph/0112106.

\bibitem{sw02} B. Schumacher and M.D. Westmoreland, quant-ph/0201061.

\bibitem{bw} C.H. Bennett and S.J. Wiesner, Phys. Rev. Lett. {\bf 69}, 2881 (1992).

\bibitem{be} A. Barenco and A. Ekert, J. Mod. Opt. {\bf 42}, 1253 (1995).

\bibitem{hjsww} P. Hausladen, R. Jozsa, B. Schumacher, M. Westmoreland and W.K. Wootters,
Phys. Rev. A {\bf 54}, 1869 (1996).

\bibitem{mro} S. Mozes, J. Oppenheim and B. Reznik, Phys. Rev. A {\bf 71}, 012311 (2005); quant-ph/0403189.

\bibitem{pati} A.K. Pati, P. Parashar and P. Agrawal, quant-ph/0412039.

\bibitem{wcsg} S. Wu, S.M. Cohen, Y. Sun and R.B. Griffiths, Phys. Rev. A {\bf 73}, 042311 (2006); quant-ph/0512169.

\bibitem{chefles} A. Chefles Phys. lett. A {\bf 239}, 339-347 (1998).
\bibitem{ivanovic}I.D. Ivanovic, Phys. Lett. A {\bf 123}, 257 (1987).
\bibitem{dieks}D. Dieks, Phys. Lett. A {\bf 126}, 303 (1988).
\bibitem{peres}A. Peres, Phys. Lett. A {\bf 128}, 19 (1988).
\bibitem{jaeger}G. Jaeger and A. Shimony, Phys.  Lett.  A {\bf 197}, 83 (1995).
\bibitem{duan}L.M. Duan and G.C. Guo, Phys. Rev. Lett. {\bf 80}, 4999 (1998).
\bibitem{chefles2}A. Chefles and S.M. Barnett, Phys. Lett. A {\bf 250}, 223 (1998).
\bibitem{sun1}Y. Sun, M. Hillery and J.A. Bergou, Phys. Rev. A {\bf 64}, 022311 (2001).

\bibitem{equeigen}
If $M$ is a square matrix, then the two matrices, $M^{\dag} M$ and $M M^{\dag}$, have the same
set of eigenvalues.

\end{thebibliography}
}
\end{document}